\documentclass[a4paper,11pt]{article}
\usepackage{pos}
\usepackage{gensymb}

\title{Matched Runs Method to Study Extended Regions of Gamma-ray Emission}
 \ShortTitle{MRM to Study Extended Regions of Gamma-ray Emission}

\author*[a]{Binita Hona}
\forColl{VERITAS}

\affiliation[a]{University of Utah \\
  Salt Lake City, UT, USA}

\emailAdd{binita.hona@utah.edu}

\abstract{Imaging atmospheric Cherenkov telescopes, such as the Very Energetic Radiation Imaging Telescope Array System (VERITAS), are uniquely suited to resolve the detailed morphology of extended regions of gamma-ray emission. However, standard VERITAS data analysis techniques have insufficient sensitivity to gamma-ray sources spanning the VERITAS field of view (3.5°), due to difficulties with background estimation. For analysis of such spatially extended sources with 0.5° to greater than 2° radius, we developed the Matched Runs Method. This method derives background estimations for observations of extended sources using matched separate observations of known point sources taken under similar observing conditions. Our technique has been validated by application to archival VERITAS data. Here we present a summary of the Matched Runs Method and multiple validation studies on different gamma-ray sources using VERITAS data.}

\FullConference{37$^{\rm{th}}$ International Cosmic Ray Conference (ICRC 2021)\\
		July 12th -- 23rd, 2021\\
		Online -- Berlin, Germany}

\begin{document}
\maketitle

\section{Introduction}
The Very Energetic Radiation Imaging Telescope Array System (VERITAS) is an array of 4 Imaging Atmospheric Cherenkov Telescopes (IACTs) located at an altitude 1268 m in southern Arizona \cite{veritas2006}. Each telescope consists of 345 hexagonal facets and 499 high quantum efficiency photomultiplier tubes \cite{Kieda:2013wom} covering a 3.5\degree ~field of view. It is sensitive to gamma rays in the energy range of 100 GeV to > 30 TeV and can detect gamma rays from a 10 mCrab point-source at a significance of $5\sigma$ in less than 25 hours of observation. However, the sensitivity of VERITAS's standard point source analysis techniques significantly decreases as the angular extension of the source increases. Therefore, the observation of extended regions of gamma-ray emission with VERITAS is limited and there has been no conclusive detection of the gamma-ray sources with large angular extent such as Geminga PWN \cite{geminga}, Cygnus Cocoon \cite{cocoon}, and 2HWC J2006+341 \cite{2017ApJ...843...40A} etc. 

The Matched Runs Method (MRM) is developed as a background analysis technique to increase VERITAS sensitivity to extended sources \cite{mrm2016}. This method estimates background using archived observations on the known point sources after carefully matching the observing conditions such as elevation, azimuth angle, weather quality, date of the observation and cosmic-ray rate with the region of interest. We performed a series of validation tests of MRM using the Eventdisplay package \cite{eventdisplay}. The method has been successfully validated on Segue 1, Ursa Minor \cite{mrmicrc2019} and weak blazars like 1ES 0229+200. Here, we present the results of the validation tests on the highly variable bright blazar Mrk 421 and the mildly extended source IC 443.

\section{Background Estimation with MRM Method}
There are two methods used for background calculations in the VERITAS standard analysis. They are known as the Ring Background (RB) and the Reflected Region (RE) methods \cite{stbkg}. In the RB method, the background is estimated from  a ring around the gamma-ray emission or ON region \cite{veritas2008}. In the Reflected Region method, the background is estimated from a number of regions, of the same size as the ON region, placed in a ring arrangement offset from the camera center \cite{veritas2008}. Both of these background estimation techniques require large regions in the camera’s field of view with no source emission. Hence, when the angular extent of the source becomes close to or greater than the field of view, there is not sufficient area remaining to use for the background calculation. Hence, both RB and RE methods are not sensitive to extended regions (>0.5\degree spatial extent) of gamma-ray emission.

An alternative technique could be applying ON/OFF method in which the observation on a region of interest is ON observation and away from the region of interest is OFF observation. This way, the camera can utilize the entire field of view for the background observation. The OFF observations are taken on the same day in a blank sky patch with no known gamma-ray sources under similar observing parameters such as elevation and azimuth. The classical ON/OFF technique has been successfully used by gamma-ray observatories to observe sources with a spatial extent comparable to the field of view of the camera. However, this technique requires twice as much observation time compared to the standard VERITAS background estimation methods (RB and RE).

As a better alternative to the classical ON/OFF method, the matched run method was developed \cite{mrmicrc2019}. Instead of observing a dedicated OFF run for each ON run, it utilizes archived observations taken on a  known point source under similar observing conditions and with similar background rates. The run from the archived data which matches the background rate of an ON run is called a "matched run" \cite{mrmicrc2019}. Since the matched runs are observations taken on a known source, we then have to apply the standard background techniques (RB or RE) to estimate the background.  Hence, the MRM is a combination of the classical ON/OFF method and the VERITAS standard background estimation technique.

The MRM match finding algorithm takes into account four parameters: elevation angle, azimuth angle, MJD of the run and number of non-gamma or cosmic ray events $N^{CR}$. A user supplies the maximum difference allowed for these four parameters between an ON run and an OFF run along with a list of ON observation runs and a list OFF observation runs. Based on the list of OFF observation runs provided, the match finding algorithm compiles a list of matched runs with minimum difference in $N^{CR}$, $\Delta$, where

\begin{equation}
\Delta = \sum_{i=0}^{n} N^{CR}_{i_{ON}} - N^{CR}_{i_{Matched}}.   
\end{equation}
$N^{CR}_{i_{ON}}$ is the $N^{CR}$ of the $i^{th}$ ON run, $N^{CR}_{i_{Matched}}$ is the $N^{CR}$ of its matched run, and $n$ is the number of ON runs. 

The algorithm will continue to provide a list of matched runs until the number of OFF observation runs is less than the number of ON observation runs.  For each matched list the algorithm also calculates the parameter $M$, where 

\begin{equation}
    M = \frac{\sum_{i=1}^{n} N^{CR}_{i_{ON}} - N^{CR}_{i_{Matched}}}{\sqrt{\frac{\sum_{i=1}^{n} N^{CR}_{i_{ON}} + N^{CR}_{i_{Matched}}}{n}}}.
\end{equation}

$M$ determines how well the ON run list matches with the matched run list. The matched run list with minimum $M$ is selected as the best set of matched runs.

\section{Validation Tests}
We performed a series of tests on archival VERITAS observations using MRM with the EventDisplay package. These validation tests include analysing the various source types detected by VERITAS using the standard background estimation technique and comparing their detection significance with the results obtained with the MRM background calculation. The data collected from 2009 through summer 2012 are categorized as V5 epoch and the data collected after 2012 are categorized as V6 epoch \cite{Kieda:2013wom}. An energy threshold of 450 GeV is used. 

\begin{figure}
\includegraphics[width=0.5\textwidth]{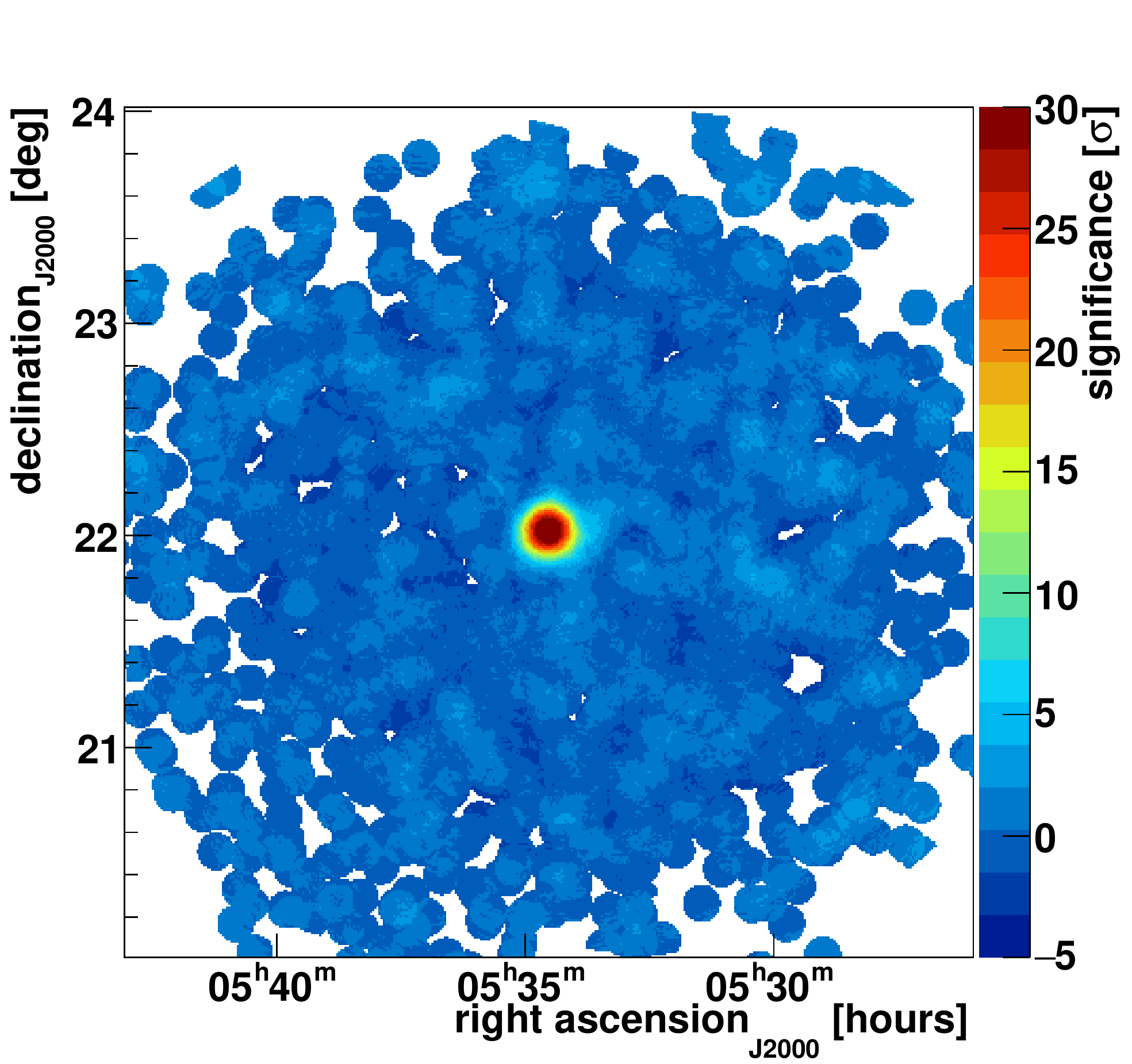}
\includegraphics[width=0.5\textwidth]{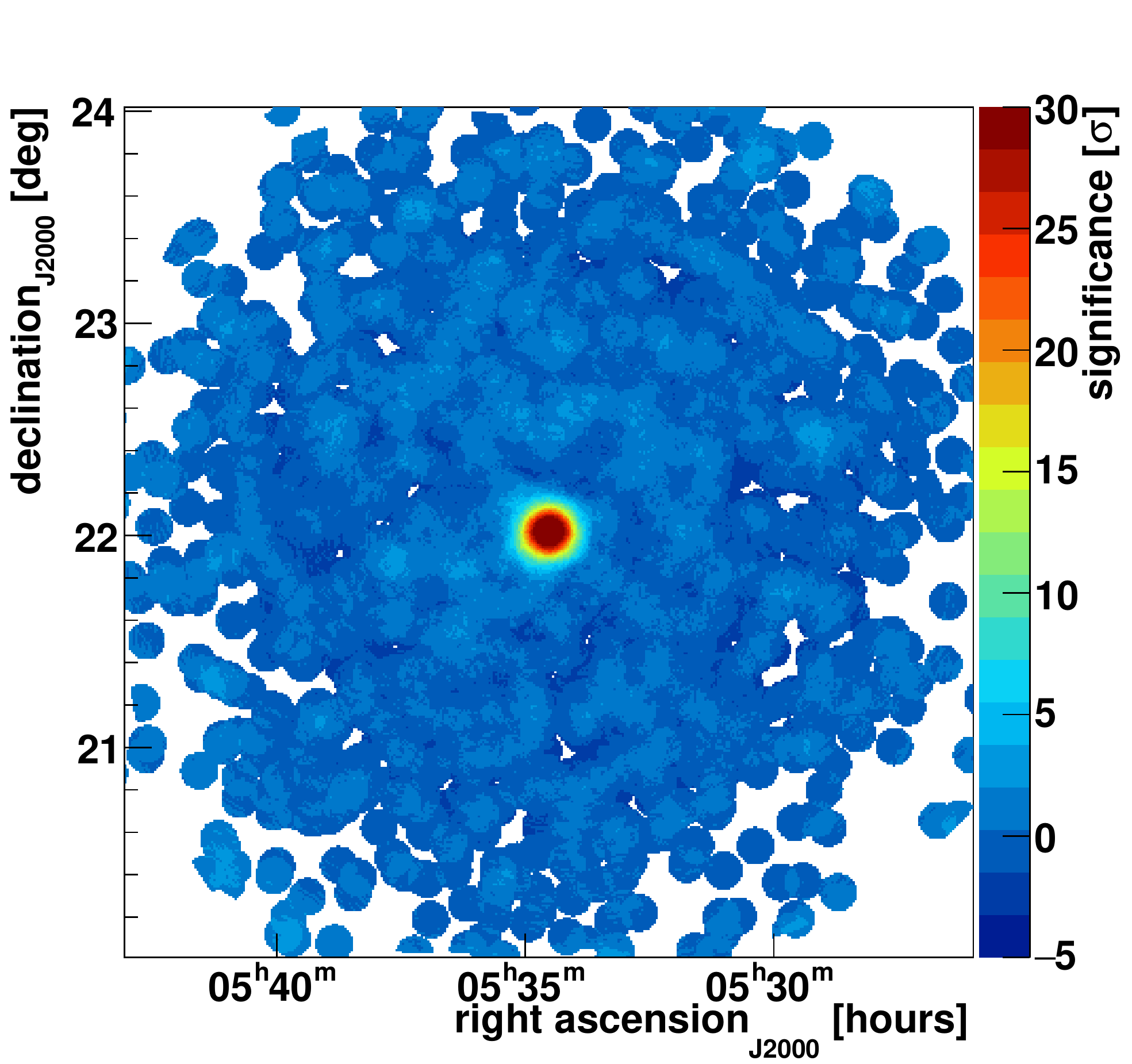}
\caption{ \label{fig:crab}
\textbf{Left}: Significance map of Crab nebula using standard analysis.  
\textbf{Right}: Significance map of Crab nebula using MRM analysis.}
%\end{minipage}
\end{figure}

\begin{figure}
\includegraphics[width=0.5\textwidth]{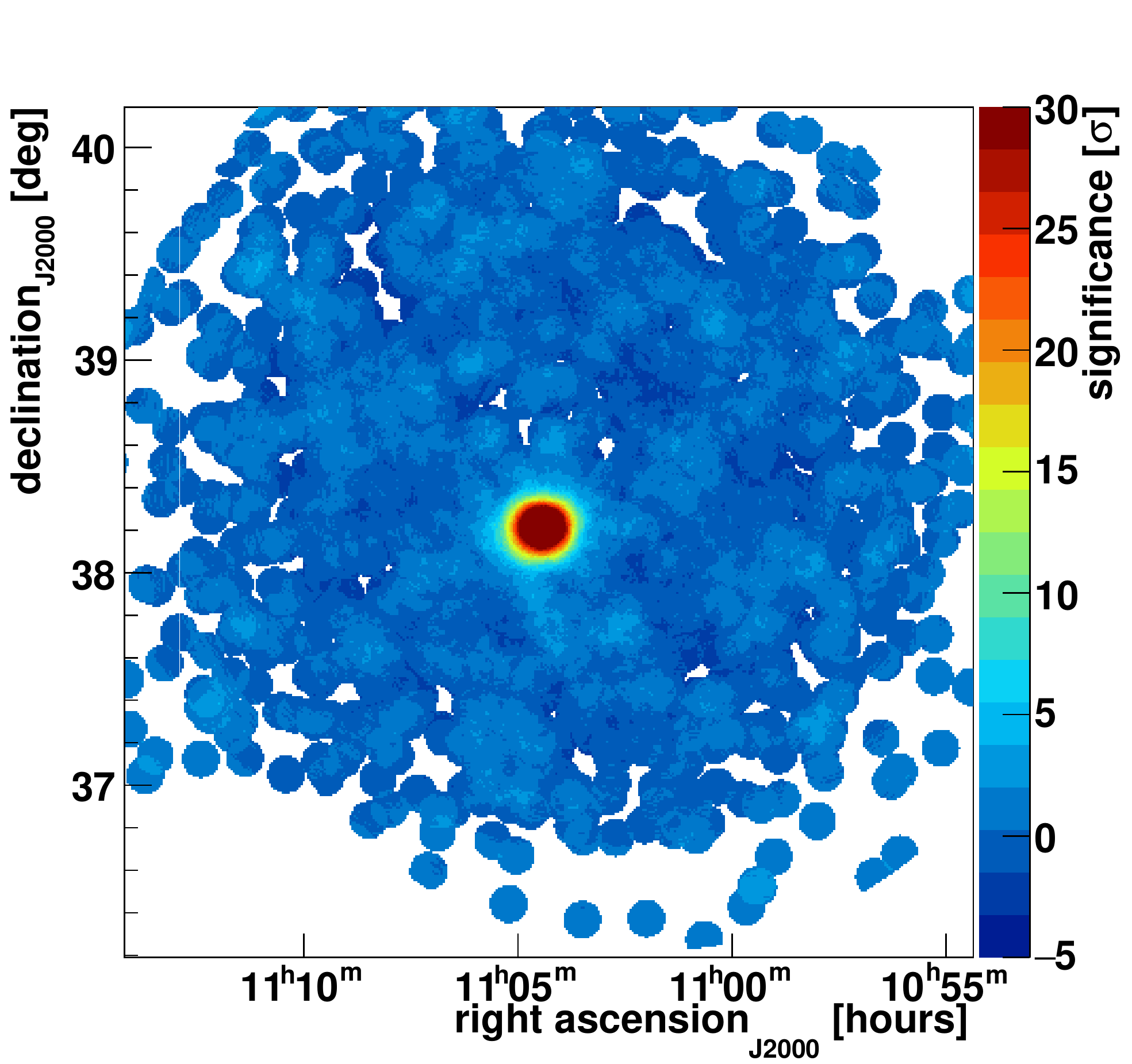}
\includegraphics[width=0.5\textwidth]{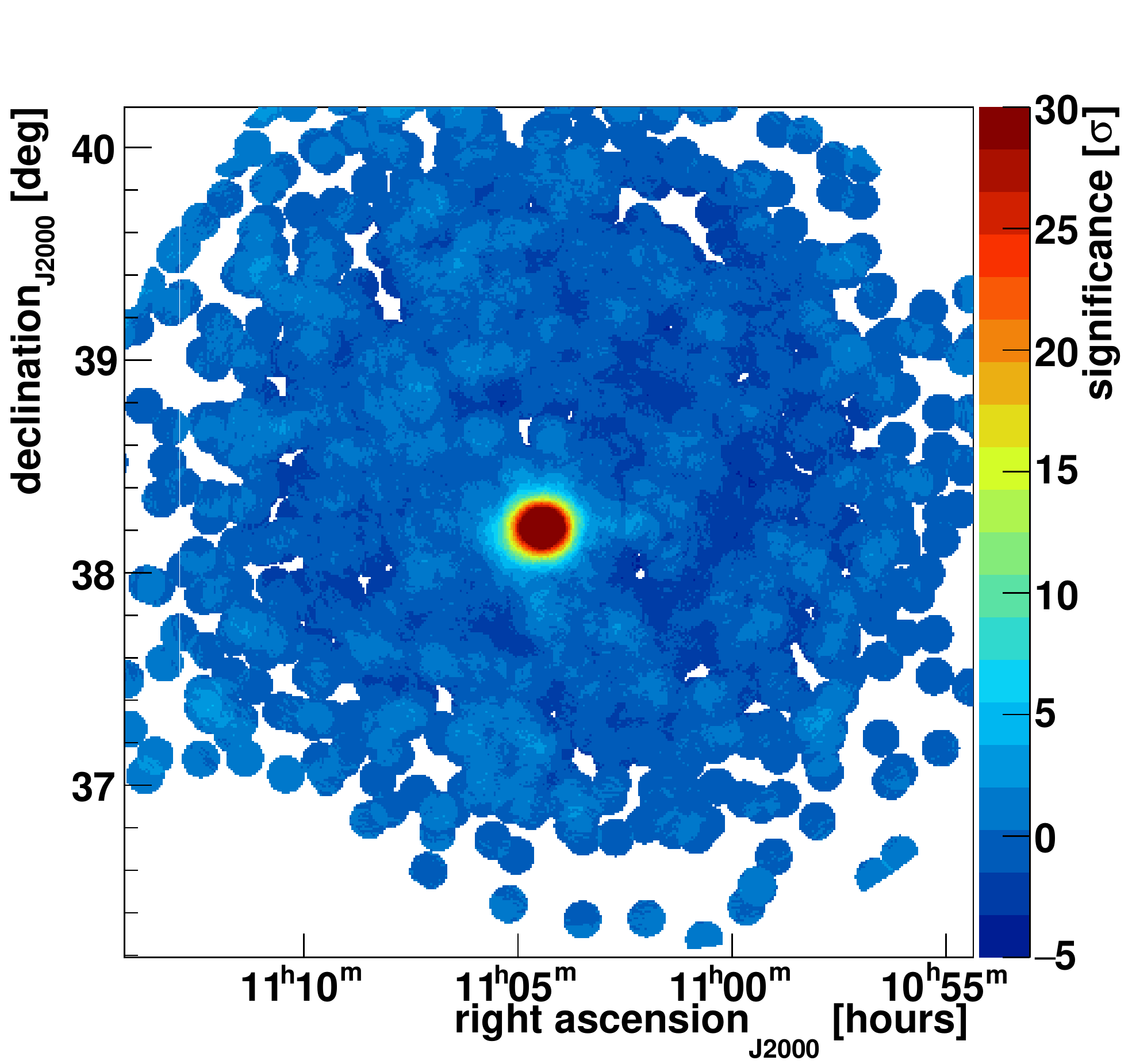}
\caption{ \label{fig:bright}
\textbf{Left}: Significance map of MRK 421 flaring state using standard analysis.  
\textbf{Right}: Significance map of MRK 421 flaring state using MRM analysis.}
%\end{minipage}
\end{figure}

\begin{figure}
\includegraphics[width=0.5\textwidth]{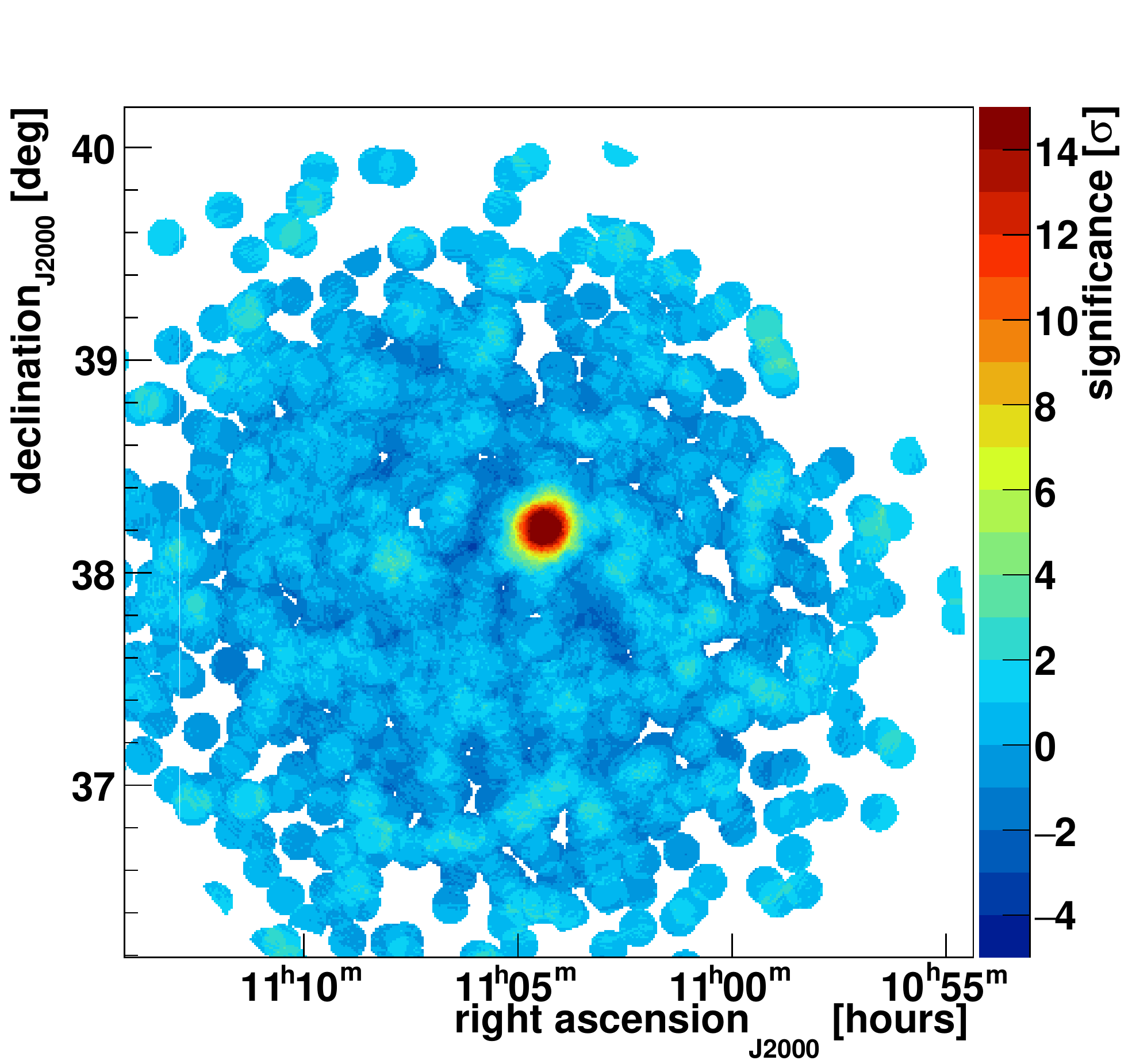}
\includegraphics[width=0.5\textwidth]{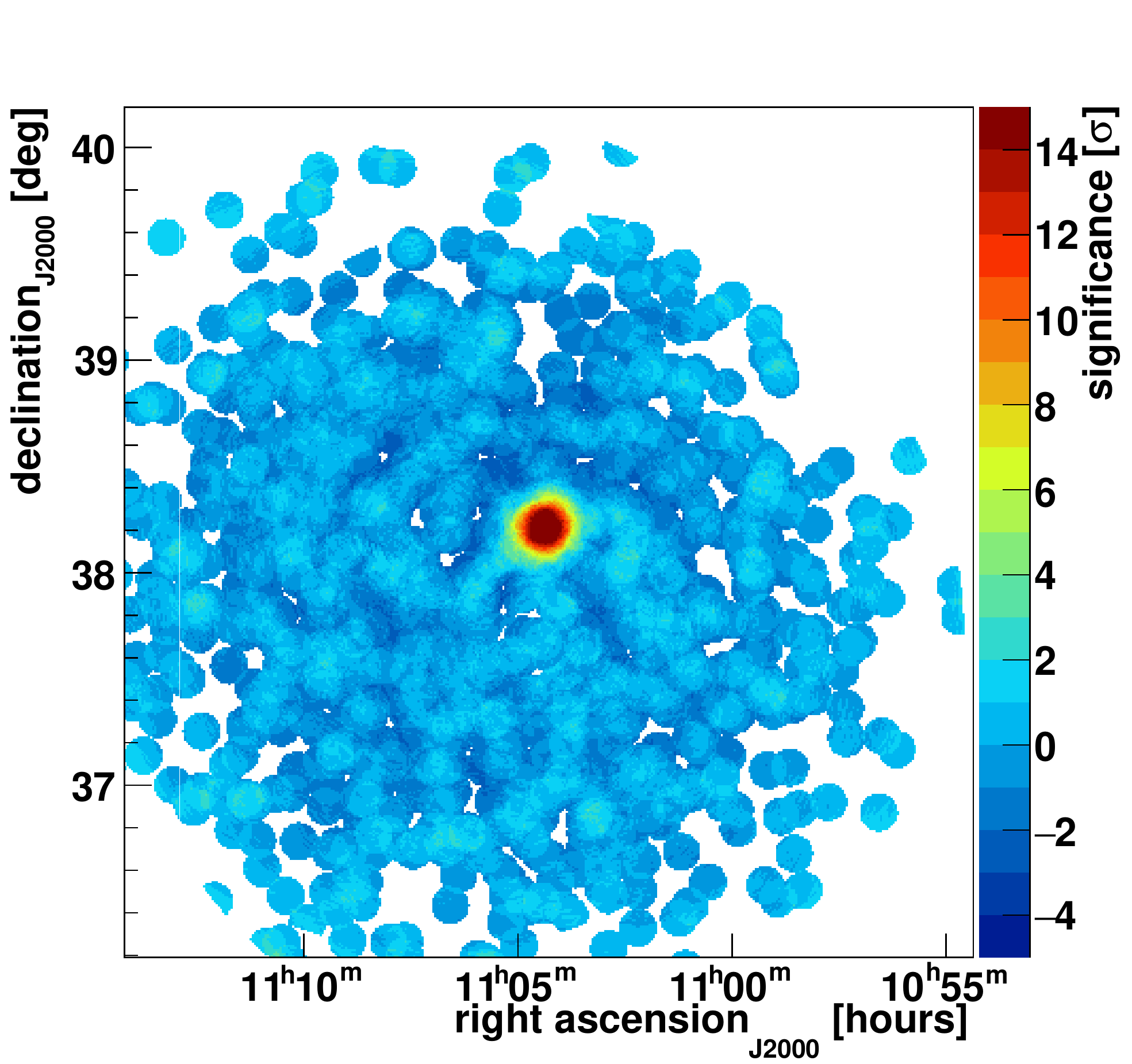}
\caption{ \label{fig:qui}
\textbf{Left}: Significance map of MRK 421 quiescent state using standard analysis.  
\textbf{Right}: Significance map of MRK 421 quiescent state using MRM analysis.}
%\end{minipage}
\end{figure}

\begin{figure}
\includegraphics[width=0.5\textwidth]{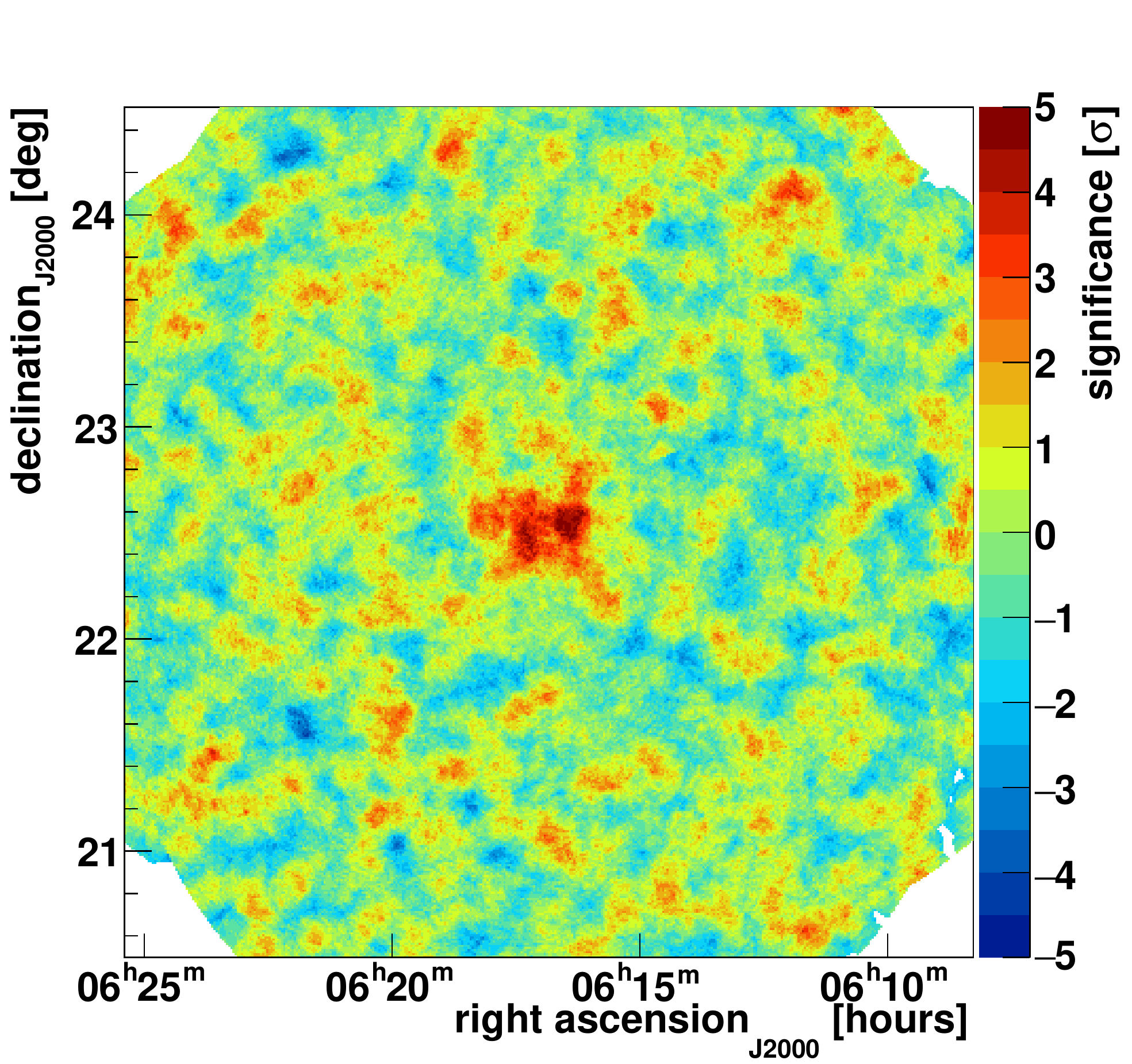}
\includegraphics[width=0.5\textwidth]{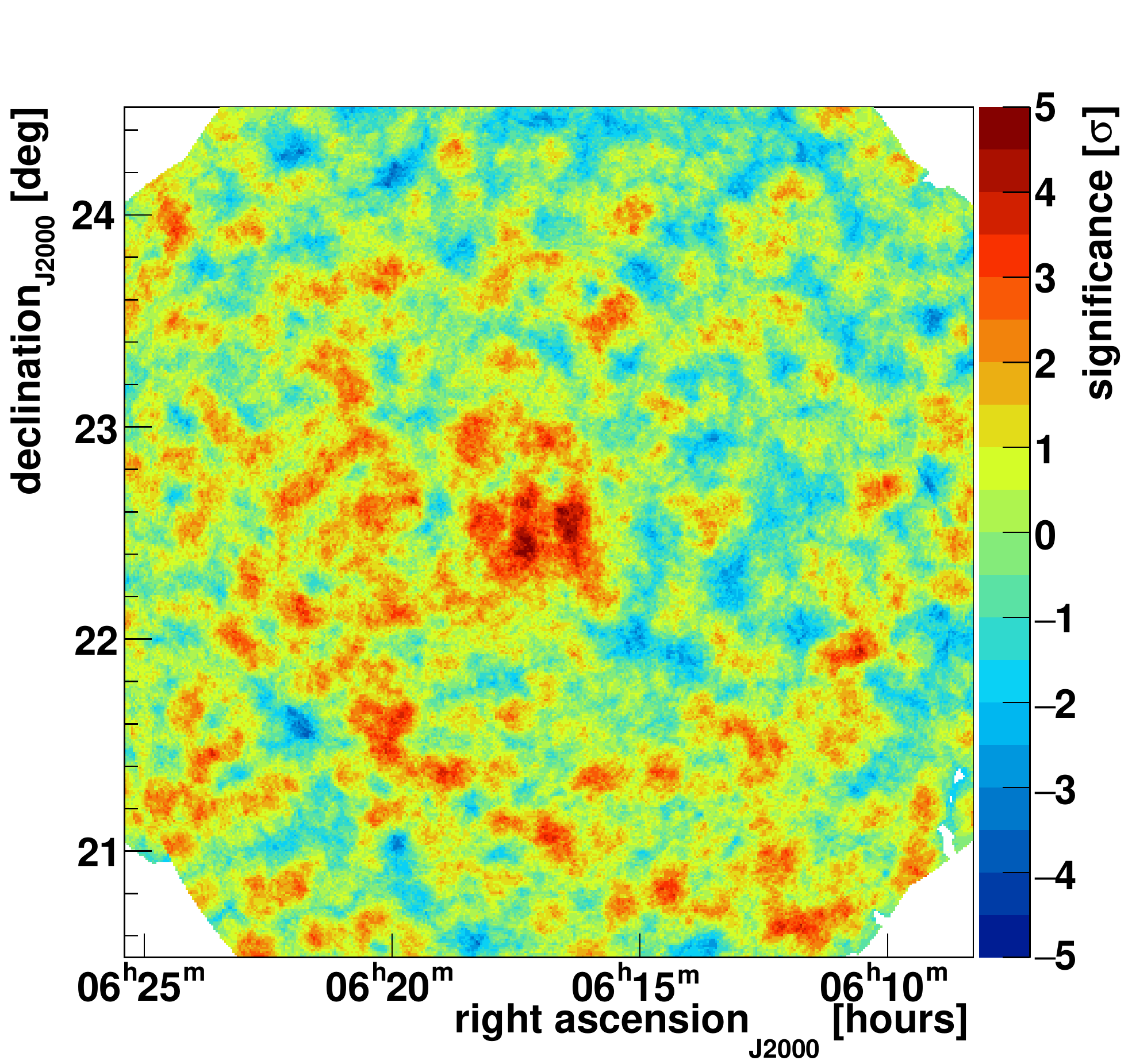}
\caption{ \label{fig:ic443}
\textbf{Left}: Significance map of IC 443  using standard analysis. 
\textbf{Right}: Significance map of IC 443 using MRM analysis.}
%\end{minipage}
\end{figure}

Fig. \ref{fig:crab} shows the validation test on 2 hr V6 epoch Crab data. Using the standard RE method for the background estimation, the Crab is detected at a significance of 35.1 $\sigma$ and using MRM for the background estimation, the source is detected at a significance of 34.5 $\sigma$. From our studies on the Crab nebula, there is about 10 \% systematic uncertainty between the two background estimation methods.
Fig. \ref{fig:bright} shows the results for the blazar MRK 421 in a bright state using 2.5 hr V6 epoch data. Using the standard RE method, MRK 421 is detected at a significance of 46.9 $\sigma$ and using the MRM  background estimation, the source is detected at a significance of 45.6 $\sigma$. Fig. \ref{fig:qui} shows the results for MRK 421 at quiescent state using 2 hr of V6 epoch data. Using standard background estimation, the blazar was detected at a significance of 18.5 $\sigma$ and using MRM, it is detected at a significance of 17.9 $\sigma$. These results validate that the MRM is sensitive to the highly variable blazar. Fig. \ref{fig:ic443} shows the results for a slightly extended source IC 443 with 0.16\degree spatial extent, using 22 hr V5 epoch data. It is detected at a significance of 3.0 $\sigma$ using standard analysis and using MRM, it is detected at a significance of 2.8 $\sigma$. 

\section{Summary}
In all the validation tests, the MRM is able to recover the significances obtained with the VERITAS standard analysis. For IC 443, while the two methods agree, using 22 hr of data, the detection was only about 3 $\sigma$. We are in the process of analyzing both V5 and V6 epoch data thereby increasing our data set to increase the detection significance. We will also repeat the validation studies at an energy threshold of 250-270 GeV. Once the validation tests are completed, the MRM will be applied to analyze the Geminga \cite{geminga} and Cygnus Cocoon \cite{cocoon} region data.

\section{Acknowledgement}
This research is supported by grants from the U.S. Department of Energy Office of Science, the U.S. National Science Foundation and the Smithsonian Institution, by NSERC in Canada, and by the Helmholtz Association in Germany. This research used resources provided by the Open Science Grid, which is supported by the National Science Foundation and the U.S. Department of Energy's Office of Science, and resources of the National Energy Research Scientific Computing Center (NERSC), a U.S. Department of Energy Office of Science User Facility operated under Contract No. DE-AC02-05CH11231. We acknowledge the excellent work of the technical support staff at the Fred Lawrence Whipple Observatory and at the collaborating institutions in the construction and operation of the instrument.

\clearpage
\bibliographystyle{plain}
\bibliography{bibliography}

%\begin{thebibliography}{99}
%\bibitem{...}

%\end{thebibliography}

\clearpage \section*{Full Authors List: \Coll\ Collaboration}

\scriptsize
\noindent
C.~B.~Adams$^{1}$,
A.~Archer$^{2}$
W.~Benbow$^{3}$,
A.~Brill$^{1}$,
J.~H.~Buckley$^{4}$,
M.~Capasso$^{5}$,
J.~L.~Christiansen$^{6}$,
A.~J.~Chromey$^{7}$, 
M.~Errando$^{4}$,
A.~Falcone$^{8}$,
K.~A.~Farrell$^{9}$,
Q.~Feng$^{5}$,
G.~M.~Foote$^{10}$,
L.~Fortson$^{11}$,
A.~Furniss$^{12}$,
A.~Gent$^{13}$,
G.~H.~Gillanders$^{14}$,
C.~Giuri$^{15}$,
O.~Gueta$^{15}$,
D.~Hanna$^{16}$,
O.~Hervet$^{17}$,
J.~Holder$^{10}$,
B.~Hona$^{18}$,
T.~B.~Humensky$^{1}$,
W.~Jin$^{19}$,
P.~Kaaret$^{20}$,
M.~Kertzman$^{2}$,
T.~K.~Kleiner$^{15}$,
S.~Kumar$^{16}$,
M.~J.~Lang$^{14}$,
M.~Lundy$^{16}$,
G.~Maier$^{15}$,
C.~E~McGrath$^{9}$,
P.~Moriarty$^{14}$,
R.~Mukherjee$^{5}$,
D.~Nieto$^{21}$,
M.~Nievas-Rosillo$^{15}$,
S.~O'Brien$^{16}$,
R.~A.~Ong$^{22}$,
A.~N.~Otte$^{13}$,
S.~R. Patel$^{15}$,
S.~Patel$^{20}$,
K.~Pfrang$^{15}$,
M.~Pohl$^{23,15}$,
R.~R.~Prado$^{15}$,
E.~Pueschel$^{15}$,
J.~Quinn$^{9}$,
K.~Ragan$^{16}$,
P.~T.~Reynolds$^{24}$,
D.~Ribeiro$^{1}$,
E.~Roache$^{3}$,
J.~L.~Ryan$^{22}$,
I.~Sadeh$^{15}$,
M.~Santander$^{19}$,
G.~H.~Sembroski$^{25}$,
R.~Shang$^{22}$,
D.~Tak$^{15}$,
V.~V.~Vassiliev$^{22}$,
A.~Weinstein$^{7}$,
D.~A.~Williams$^{17}$,
and 
T.~J.~Williamson$^{10}$\\
\noindent
$^1${Physics Department, Columbia University, New York, NY 10027, USA}
$^{2}${Department of Physics and Astronomy, DePauw University, Greencastle, IN 46135-0037, USA}
$^3${Center for Astrophysics $|$ Harvard \& Smithsonian, Cambridge, MA 02138, USA}
$^4${Department of Physics, Washington University, St. Louis, MO 63130, USA}
$^5${Department of Physics and Astronomy, Barnard College, Columbia University, NY 10027, USA}
$^6${Physics Department, California Polytechnic State University, San Luis Obispo, CA 94307, USA} 
$^7${Department of Physics and Astronomy, Iowa State University, Ames, IA 50011, USA}
$^8${Department of Astronomy and Astrophysics, 525 Davey Lab, Pennsylvania State University, University Park, PA 16802, USA}
$^9${School of Physics, University College Dublin, Belfield, Dublin 4, Ireland}
$^10${Department of Physics and Astronomy and the Bartol Research Institute, University of Delaware, Newark, DE 19716, USA}
$^{11}${School of Physics and Astronomy, University of Minnesota, Minneapolis, MN 55455, USA}
$^{12}${Department of Physics, California State University - East Bay, Hayward, CA 94542, USA}
$^{13}${School of Physics and Center for Relativistic Astrophysics, Georgia Institute of Technology, 837 State Street NW, Atlanta, GA 30332-0430}
$^{14}${School of Physics, National University of Ireland Galway, University Road, Galway, Ireland}
$^{15}${DESY, Platanenallee 6, 15738 Zeuthen, Germany}
$^{16}${Physics Department, McGill University, Montreal, QC H3A 2T8, Canada}
$^{17}${Santa Cruz Institute for Particle Physics and Department of Physics, University of California, Santa Cruz, CA 95064, USA}
$^{18}${Department of Physics and Astronomy, University of Utah, Salt Lake City, UT 84112, USA}
$^{19}${Department of Physics and Astronomy, University of Alabama, Tuscaloosa, AL 35487, USA}
$^{20}${Department of Physics and Astronomy, University of Iowa, Van Allen Hall, Iowa City, IA 52242, USA}
$^{21}${Institute of Particle and Cosmos Physics, Universidad Complutense de Madrid, 28040 Madrid, Spain}
$^{22}${Department of Physics and Astronomy, University of California, Los Angeles, CA 90095, USA}
$^{23}${Institute of Physics and Astronomy, University of Potsdam, 14476 Potsdam-Golm, Germany}
$^{24}${Department of Physical Sciences, Munster Technological University, Bishopstown, Cork, T12 P928, Ireland}
$^{25}${Department of Physics and Astronomy, Purdue University, West Lafayette, IN 47907, USA}

%% Full authors list (ONLY FOR COLLABORATIONS)
%\clearpage
%\section*{Full Authors List: \Coll\ Collaboration}
%
%\noindent \textbf{Note comment afterwards:} Collaborations have the possibility to provide an authors list in xml format which will be used while generating the DOI entries making the full authors list searchable in databases like Inspire HEP. For instructions please go to icrc2021.desy.de/proceedings or contact us under icrc2021proc@desy.de.\\
%
%\scriptsize
%\noindent
%first.author$^1$, 
%second.author$^2$, 
%third.author$^3$ % .... more names
%and 
%last.author$^{n}$ \\
%
%\noindent
%$^1$first.affiliation.
%$^2$second.affiliation. % .... more affiliation
%$^{m}$last.affiliation.

\end{document}